\documentclass[10pt]{extarticle}
\usepackage[utf8]{inputenc}
\usepackage[english]{babel}
\usepackage[unicode, pdftex, colorlinks]{hyperref}
\usepackage{url}
\usepackage{csquotes}
\usepackage{graphicx}
\DeclareGraphicsExtensions{.pdf,.png,.jpg}
\usepackage[left=22mm, top=13mm, right=22mm, bottom=15mm, nohead, footskip=10mm]{geometry} 
\usepackage{amsmath}
\usepackage{amsfonts}
\usepackage[dvipsnames]{xcolor}
\usepackage{cancel}
\usepackage[normalem]{ulem}
\usepackage{colortbl}
\usepackage{booktabs}
\usepackage{array}
\usepackage[
    backend=bibtex, 
    natbib=true,
    style=phys,
    biblabel=brackets,
    giveninits=true,
    abbreviate=false,
    doi=false, 
    url=true,
    isbn=false,
    block=space,
    backref=false, 
]{biblatex}
\DeclareFieldFormat
  [article,inbook,incollection,inproceedings,patent,thesis,unpublished]
  {title}{#1\isdot}
\DeclareFieldFormat[article,inproceedings,incollection]{journaltitle}{\textit{#1}}
\begin{document}
\title{Cryptocurrencies in the Quantum Age: Migration Paths to PQC}

\author{
A. Kodukhov\\[0.5em]
\textit{Terra Quantum AG, Kornhausstrasse 25, 9000 St.\,Gallen, Switzerland}\\[0.75em]
\small Reviewed by C. Mansell, T. Hartley, R. Brasher, and E. Wood
}
\date{\today}
\maketitle

\section{Introduction}

The emergence of quantum computing brings a significant challenge to modern cryptography by introducing fundamentally new computational algorithms that can achieve dramatic speedups over the best known classical methods\,\cite{Qday_TQ}.
In particular, Shor’s algorithm threatens widely deployed public-key cryptosystems, including RSA and elliptic-curve cryptography (ECC), which form the security foundation of most blockchain networks.
ECC is especially vulnerable because its relatively small key sizes make it an attractive target for quantum attacks.
Recent results published by Google Quantum AI and collaborators~\cite{google2026} further reinforce these concerns by demonstrating that the quantum resources required to break elliptic-curve cryptography using Shor’s algorithm may be an order of magnitude smaller than previously estimated, thereby validating the need for proactive migration toward post-quantum cryptographic solutions.

The implications of quantum computing for cryptocurrencies are particularly severe because all major blockchain platforms rely on elliptic-curve cryptography to secure user accounts, authorize transactions, and maintain network consensus.
A cryptographically relevant quantum computer (CRQC),  capable of executing Shor’s algorithm, could derive private keys from public keys, enabling attackers to steal funds from vulnerable accounts.
Beyond direct asset theft, quantum attacks could compromise administrative and governance keys, allowing malicious upgrades of smart contracts.
Stablecoin ecosystems are especially exposed, as the compromise of issuer or administrator keys could enable unauthorized minting, burning, freezing, or transfer of stablecoins, potentially undermining trust in some of the most widely used digital assets.

Different blockchain platforms exhibit different levels of quantum readiness.
While most cryptocurrencies rely on elliptic-curve cryptography and are therefore vulnerable to quantum attacks, some ecosystems incorporate mechanisms that provide partial protection against quantum adversaries.
Examples include hash-based address schemes in Bitcoin, which hide public keys until funds are spent, and quantum-resistant constructions such as Winternitz-based vaults developed for Solana\,\cite{buchmann2013security, blueshift}.
In this white paper, we review the blockchain components most exposed to quantum attacks, analyze the associated economic and market risks, and discuss practical migration strategies toward post-quantum security.

The remainder of this white paper is organized as follows.
In Section\,\ref{sec:quantum_attacks}, we review the technical aspects of quantum attacks and identify the blockchain components that are vulnerable to them.
In Section\,\ref{sec:economic}, we provide an overview of the associated economic and market risks, with a particular focus on major blockchain platforms such as Bitcoin, Ethereum, and Solana.
Migration strategies toward post-quantum security, as well as existing post-quantum cryptographic solutions, are discussed in Section\,\ref{sec:PQC_migration}.

\section{Quantum Threats to Blockchain Systems}\label{sec:quantum_attacks}
\subsection{Relevant quantum algorithms}

\begin{table}[h]
\centering
\caption{Impact of Quantum Algorithms on Blockchain Cryptography}
\label{algorithms}
\renewcommand{\arraystretch}{1.2}

\begin{tabular}{l|p{4cm}|p{6cm}}
\textbf{Algorithm} & \textbf{Cryptography affected} & \textbf{Security impact} \\
\hline
Shor's Algorithm &
Elliptic-curve cryptography: Ed25519, ECDSA, BLS &
Computes the private key from the public key, breaking digital signatures \\
\hline
Grover's Algorithm &
Hash functions (SHA-256) &
Provides quadratic speedup for brute-force attacks, effectively reducing hash security level \\
\end{tabular}

\end{table}
The main quantum algorithms relevant to cryptography are summarized in Table\,\ref{algorithms}.
The primary threat to modern public-key cryptography arises from Shor's algorithm, which solves the integer factorization and discrete logarithm problems in polynomial time, whereas the best known classical algorithms require sub-exponential or exponential time.
As a result, all widely deployed asymmetric cryptographic schemes, including RSA, Diffie--Hellman, ECDSA, and EdDSA, are vulnerable to sufficiently large quantum computers.
By applying Shor's algorithm to a publicly known key, an adversary can recover the corresponding private key and subsequently forge valid digital signatures.

In the context of blockchain systems, such an attack would enable an adversary to create fraudulent transactions on behalf of legitimate users and potentially compromise validator identities and consensus mechanisms, see Subsection\,\ref{subsec:vul_components} for details.
Recent advances in quantum resource estimation have further increased the urgency of migration planning.
In particular, Google's recent work demonstrates a significant reduction in the number of logical and physical qubits required to execute Shor's algorithm against elliptic-curve cryptography compared to previous estimates.
Nevertheless, estimates of the resources required for CRQCs remain highly dependent on the underlying hardware assumptions and error-correction techniques.
For a detailed discussion of current quantum resource estimates and Q-Day scenarios, we refer the reader to our companion analysis~\cite{Qday_TQ}.

Another relevant quantum algorithm is Grover’s algorithm, which provides a quadratic speedup for unstructured search problems.
Unlike Shor’s algorithm, Grover’s algorithm is not directly applicable to solving the discrete logarithm or integer factorization problems.
Instead, it can be used to accelerate brute-force searches and attacks against symmetric cryptographic primitives, such as hash functions.
For a hash function with an $n$-bit security level, Grover’s algorithm effectively reduces the security level to approximately $n$/2 bits.
For example, the preimage resistance of a 256-bit hash function\,(e.g. SHA-256) is reduced from $2^{256}$ to $2^{128}$ operations.
Despite this substantial speedup, modern cryptographic hash functions remain secure against quantum adversaries when sufficiently large output sizes are used.
Consequently, symmetric cryptography and hash-based constructions are generally considered significantly more resistant to quantum attacks than public-key cryptographic schemes.

Additional approaches to attacking cryptographic problems with quantum computers include quantum annealing and variational quantum algorithms.
Although these directions are receiving considerable research attention, they are beyond the scope of this white paper.

\subsection{Attack Scenarios}

Quantum attacks against blockchain systems may be categorized according to the required execution speed and exposure model.

\subsubsection*{At-Rest Attacks}

At-rest attacks target public keys that remain exposed for long periods of time. Examples include dormant wallets, validator keys, or reused addresses.

These attacks can be performed even by relatively slow CRQCs because the attacker may spend hours or days deriving the corresponding private key offline.

At-rest attacks represent the most immediate and realistic quantum threat to blockchain systems.

\subsubsection*{On-Spend Attacks}
On-spend attacks occur when a user broadcasts a transaction and exposes the public key during transaction propagation.
The attacker must derive the private key before the transaction is finalized on-chain.
Such attacks require fast-clock CRQCs capable of solving ECDLP problems within seconds or minutes.
The feasibility of on-spend attacks depends strongly on blockchain confirmation latency:
\begin{itemize}
\item Solana: $\sim$400 ms,
\item Ethereum: $\sim$12 s,
\item Bitcoin: $\sim$10 min.
\end{itemize}
As a result, Bitcoin is the most vulnerable to the on-spend attacks, since it has the highest confirmation latency.
Although Bitcoin addresses add a layer of hash-based protection, funds can still be stolen by fast-clock CRQCs once the public key is broadcast.

\subsubsection*{On-Setup Attacks}

Certain cryptographic constructions may permit one-time quantum attacks that produce reusable backdoors into cryptographic protocols. These attacks are particularly relevant for:
\begin{itemize}
\item zero-knowledge proof systems,
\item bridge infrastructures,
\item data availability sampling,
\item advanced privacy protocols.
\end{itemize}
Although less discussed publicly, such attacks could have catastrophic systemic consequences.

\subsection{Vulnerable blockchain components}\label{subsec:vul_components}
In Bitcoin, funds are stored as unspent transaction outputs (UTXOs) rather than account balances.
In many common address types, the address is derived from a hash of the public key used to authorize spending.
As a result, funds remain protected from at-rest quantum attacks as long as the corresponding public key has not been revealed.
Once a transaction is broadcast, however, the public key becomes visible, creating a window for an on-spend attack.
Because Bitcoin’s average block time is approximately 10 minutes, this window is relatively long compared with high-throughput blockchains.
Addresses whose public keys have already been exposed through previous transactions, address reuse, or off-chain disclosure are additionally vulnerable to at-rest attacks.
By contrast, Bitcoin’s proof-of-work mining process is not considered a near-term quantum vulnerability, since Grover’s algorithm provides only a quadratic speedup and classical mining hardware remains highly optimized.

Ethereum has a different quantum-risk profile.
Its shorter block time reduces the feasibility of on-spend attacks against ordinary transactions.
However, Ethereum still relies on ECC at several critical layers.
User accounts depend on ECDSA signatures, while proof-of-stake validators rely on aggregated BLS signatures.
A quantum adversary capable of deriving private keys from public keys could therefore compromise user accounts, validator identities, and consensus-related infrastructure.

Solana’s exposure differs from Bitcoin’s because its externally owned accounts are Ed25519 public keys.
Unlike Bitcoin address types that hide the public key behind a hash until spending, Solana accounts reveal the public key from the creation of an account.
Therefore, once a CRQC becomes available, externally controlled Solana accounts become vulnerable without requiring any prior outgoing transaction\,\cite{anza, blueshift, jumpcrypto}.
At the same time, Solana’s very short block time makes on-spend attacks less practical than in Bitcoin.
The more immediate concern is at-rest exposure of accounts, validator keys, and signing authorities.

Not all Solana addresses have the same risk profile.
Program Derived Addresses (PDA) are generated through hash-based mechanisms and do not correspond to ordinary Ed25519 private keys.
They are therefore comparatively resistant to Shor-type attacks.
Nevertheless, many PDAs are controlled by upgrade authorities, multisig signers, or other externally owned accounts.
If those controlling keys remain Ed25519-based, the surrounding application infrastructure may still be vulnerable even when the PDA itself is not.

Smart-contract ecosystems introduce additional high-value attack surfaces.
In Ethereum, Solana, and other programmable blockchains, many critical operations are controlled by privileged keys rather than by the base protocol alone.
Quantum compromise of such keys could enable unauthorized minting or burning of stablecoins, freezing or seizure of tokenized assets, malicious smart-contract upgrades, compromise of cross-chain bridges, or manipulation of decentralized governance systems.
These privileged accounts, including multisigs, DAO treasuries, stablecoin administrator keys, bridge operators, and upgrade authorities, are likely to be among the first targets of a quantum attacker.

Overall, the most vulnerable blockchain components are not limited to ordinary user wallets.
The highest-risk targets include exposed public-key accounts, reused or previously spent addresses, validator and consensus keys, bridge operators, stablecoin administrator keys, smart-contract upgrade authorities, DAO treasuries, and custodial hot wallets.
These components combine cryptographic exposure with high economic value, making them priority targets in a future quantum attack scenario.

\section{Economic and Market Risk Analysis}\label{sec:economic}

The security assumptions underlying wallets, validators, bridges, smart contracts, and custodial infrastructure may eventually become invalid. 
Cryptocurrencies are particularly vulnerable compared to traditional financial systems for two principal reasons. First, blockchain systems prioritize efficiency and scalability, frequently employing compact elliptic-curve schemes with relatively small key sizes. Second, blockchain transactions are generally irreversible and pseudonymous, meaning that a forged signature immediately enables irreversible theft without institutional recovery mechanisms\,\cite{google2026}.

The economic exposure is already enormous and is expected to increase significantly over the coming decade due to the rapid growth of tokenized assets, decentralized finance (DeFi), and fiat-backed stablecoins. According to recent projections, tokenization of real-world assets (RWAs) and stablecoins may increase the value of blockchain-governed assets by nearly an order of magnitude by 2030.

Among existing blockchain ecosystems, Bitcoin, Ethereum, Solana, and stablecoin infrastructures represent the largest concentration of quantum-vulnerable assets.

\subsection{Bitcoin Exposure}

Bitcoin\,\cite{nakamoto2008bitcoin} exhibits a varying quantum security level depending on the script type used to secure funds. 
In the classical Pay-to-Public-Key (P2PK) scheme, public keys are directly exposed on-chain and therefore remain continuously vulnerable to at-rest quantum attacks.
Approximately 1.7 million BTC (nearly 9\% of all Bitcoin) are currently protected by P2PK scripts\,\cite{google2026}.

More modern script types such as Pay-to-Public-Key-Hash (P2PKH) partially mitigate this exposure by hiding the public key behind a hash until the funds are spent.
Nevertheless, once a transaction is broadcast, the public key becomes visible, opening a window for an on-spend attack.

The total amount of Bitcoin considered quantum-vulnerable is estimated at approximately 6.9 million BTC\,\cite{google2026}, corresponding to nearly 500 billion USD at current market valuations.
This estimate includes funds held in script types that are vulnerable to at-rest attacks (see Table~\ref{tab:btc_quantum_exposure}), as well as funds held in script types that are resistant to at-rest attacks but whose public keys have already been exposed.
Dormant wallets, including lost or abandoned coins, represent a particularly severe problem because their owners cannot migrate them to post-quantum schemes.

\begin{table}[ht]
\centering
\caption{Quantum Exposure of Major Bitcoin Script Types. The table is generated using data from the Google Cloud BigQuery public Bitcoin dataset~\cite{google_bigquery_bitcoin} (July 2026).}
\label{tab:btc_quantum_exposure}
\renewcommand{\arraystretch}{1.2}
\begin{tabular}{lrcl}
\toprule
\textbf{Script Type} &
\textbf{Current Balance} &
\textbf{Share} &
\textbf{Quantum Attack Primary Vector} \\
\midrule
P2PK                & 1.72 M BTC & 8.6\% & At-rest \& on-spend \\
P2MS                & 40.6 BTC    & 0.0002\% & At-rest \& on-spend \\
P2TR                & 0.23 M BTC & 1.1\% & At-rest \& on-spend \\
P2PKH               & 4.58 M BTC & 22.8\% & On-spend (at-rest if address reuse) \\
P2SH                & 3.92 M BTC & 19.5\% & On-spend (at-rest if address reuse) \\
P2WPKH              & 8.19 M BTC & 40.8\% & On-spend (at-rest if address reuse)\\
P2WSH               & 1.43 M BTC & 7.1\% & On-spend (at-rest if address reuse) \\
\bottomrule
\end{tabular}
\end{table}

Dormant assets create a unique governance and economic challenge. Unlike active wallets, they cannot be upgraded through standard migration procedures.
Consequently, hundreds of billions of dollars may eventually become accessible to quantum attackers.
This may force blockchain communities to make unprecedented decisions regarding frozen assets, forced migrations, or protocol-level interventions.

\subsection{Ethereum and Smart-Contract Ecosystems}

Ethereum\,\cite{buterin2013ethereum, wood2014ethereum} and EVM-compatible systems exhibit a different quantum threat profile from Bitcoin.
Ethereum accounts rely on ECDSA signatures, while validators depend on BLS signatures for consensus participation.
Smart contracts additionally introduce new attack surfaces related to governance keys, bridges, oracles, and upgrade mechanisms.

Current (July 2026) estimates suggest that vulnerable Ethereum assets include:
\begin{itemize}

\item approximately 21.4 million ETH held by externally owned accounts among the 1,000 largest accounts, whose public keys have previously been revealed, based on the Google Cloud BigQuery public Ethereum dataset~\cite{google_bigquery_ethereum};

\item approximately 41 million ETH securing Ethereum's proof-of-stake consensus, as reported by Ultrasound Money~\cite{ultrasound_money};

\item more than 200 billion USD in stablecoins and tokenized real-world assets issued on Ethereum~\cite{rwa_xyz};

\item at least 15 million ETH in total value secured (TVS) across major quantum-vulnerable Layer-2 protocols and cross-chain bridges, based on L2BEAT data~\cite{l2beat_tvs}.

\end{itemize}

\subsection{Solana and High-Performance Chains}

The quantum threat model for Solana is particularly severe because every Solana address directly corresponds to an Ed25519 public key. Unlike Bitcoin’s UTXO model, Solana does not hide public keys behind hashes prior to spending.
As a result, all Solana accounts become immediately vulnerable once a sufficiently powerful CRQC capable of executing Shor’s algorithm becomes available. This vulnerability applies even to accounts that have never initiated a transaction.

At the same time, Solana’s extremely short block times (approximately 400 milliseconds) substantially complicate on-spend attacks, meaning that the earliest quantum attacks will most likely target dormant or continuously exposed accounts.

\subsection{Attack Prioritization}

A rational quantum attacker is unlikely to target ordinary retail wallets initially. Instead, attackers will prioritize accounts and infrastructures with the largest expected return on investment.

The most probable first targets include:

\begin{enumerate}
\item Dormant whale wallets,
\item Satoshi-era Bitcoin addresses,
\item Exchange hot wallets,
\item Stablecoin administrative keys,
\item Validator infrastructure,
\item Cross-chain bridges,
\item DAO governance wallets,
\item Custodial services and institutional treasuries.
\end{enumerate}

Bridges and stablecoin infrastructures are particularly attractive because they frequently secure billions of dollars through a small number of highly privileged cryptographic keys.

\subsection{Crypto Market Effects}

The impact of successful quantum attacks would extend far beyond isolated wallet thefts. Potential systemic effects include:

\begin{itemize}
\item loss of trust in blockchain immutability,
\item liquidity crises,
\item collapse of DeFi collateral systems,
\item destabilization of stablecoins,
\item validator centralization,
\item forced hard forks,
\item regulatory intervention,
\item accelerated migration races between ecosystems.
\end{itemize}

A large-scale quantum compromise of blockchain infrastructure could therefore trigger cascading financial instability across the broader digital asset ecosystem.

Consequently, post-quantum migration should not be viewed merely as a cryptographic upgrade, but rather as a critical financial stability and infrastructure resilience initiative for the entire blockchain industry.

\subsection{Cryptocurrencies and Traditional Financial Infrastructure}
In contrast to the decentralized cryptocurrencies, traditional financial infrastructure benefits from centralized governance.
Major payment networks such as Visa, Mastercard, and SWIFT operate under coordinated administrative structures that can mandate software upgrades, revoke compromised credentials, rotate cryptographic keys, and deploy migration strategies across participating institutions.
Furthermore, fraudulent transactions can often be reversed through legal, contractual, or operational processes.

At the same time, traditional finance faces a significant migration challenge at the endpoint layer.
Hundreds of millions of payment terminals, cards, hardware security modules, ATMs, embedded devices, and consumer electronics would require software or hardware upgrades to support new cryptographic standards.
Replacing or upgrading this installed base may require many years and substantial capital expenses.

Blockchain ecosystems face the opposite tradeoff.
Protocol-level upgrades can be more difficult because they require decentralized coordination among validators, developers, custodians, exchanges, and users.
Migration decisions may involve governance disputes, hard forks, and questions regarding the treatment of inactive or lost wallets.

Nevertheless, blockchain systems possess several structural advantages.
Most users interact through software wallets that can be upgraded rapidly without replacing physical infrastructure.
Validators, exchanges, and custodians operate comparatively small numbers of cryptographic endpoints relative to global payment networks.
New PQC signature schemes can be integrated directly into protocol upgrades, smart contracts, or wallet software without requiring replacement of billions of physical devices.

As a result, blockchain systems should not necessarily be viewed as more vulnerable than traditional financial infrastructure.
Rather, they face a different risk profile: traditional finance benefits from centralized coordination but suffers from a massive endpoint migration problem, whereas blockchain systems face governance and migration challenges but can often deploy cryptographic upgrades more rapidly once consensus is achieved.

\section{Post-Quantum Migration Strategies}\label{sec:PQC_migration}
In this section, we review existing approaches for achieving post-quantum security in cryptocurrency infrastructure.
\subsection{Post-Quantum Digital Signatures}
Digital signatures are the core cryptographic primitive used to authorize blockchain transactions, validator messages, governance actions, and smart-contract administration.
Consequently, a post-quantum migration must eventually replace or complement elliptic-curve-based schemes with post-quantum digital signatures.

The most relevant NIST-selected post-quantum signature schemes are ML-DSA\,\cite{dang2024module}, formerly CRYSTALS-Dilithium; SLH-DSA\,\cite{cooper2024stateless}, based on SPHINCS+; and FN-DSA\,\cite{fouque2018falcon}, based on Falcon.
ML-DSA and SLH-DSA are finalized NIST standards, while FN-DSA/Falcon remains in the standardization process, see Table\,\ref{tab:PQC_DSA}.
For blockchain systems, the main practical constraint is not only cryptographic security, but also signature size, public-key size, verification cost, and compatibility with transaction-size limits, see Ref.\,\cite{opilka2024performance} for the performance analysis.

Among these schemes, FN-DSA is particularly attractive for bandwidth-constrained blockchains because it provides compact signatures and fast verification.
This makes it suitable for transaction authorization, smart-contract verification, and validator-related use cases. ML-DSA has stronger standardization maturity and simpler implementation characteristics, but its larger signatures and public keys make it more difficult to integrate into high-throughput blockchains without increasing transaction-size limits.
SLH-DSA provides a conservative hash-based alternative, but its large signatures make it less suitable for ordinary blockchain transactions.

\begin{table}[h]
\centering
\caption{Comparison of Post-Quantum Signature Schemes}
\label{tab:pqc-solana}
\renewcommand{\arraystretch}{1.2}
\begin{tabular}{l|c|c|l|l}
\textbf{Scheme} & \textbf{Public Key} & \textbf{Signature} & \textbf{NIST Status} & \textbf{Suitable for blockchains?} \\
\hline
Ed25519 (current) & 32 B & 64 B & FIPS 186-5  & Yes (not PQC) \\
\hline
ML-DSA-44 (Dilithium) & 1312 B & 2420 B & FIPS 204 &
No --- 2420 B signature \\
\hline
FN-DSA (Falcon-512) & 897 B & $\leq$ 666 B & FIPS 206 (Draft) &
Yes --- balance of size and security \\
\hline
SLH-DSA-128s (SPHINCS+) & 32 B & 7856 B & FIPS 205 &
No --- 7856 B signature \\
\hline
HAWK-512\,\cite{HAWK} & 1024 B & 555 B & Research &
Not production-ready \\
\end{tabular}
\label{tab:PQC_DSA}
\end{table}

\subsection{Winternitz Vaults on Solana}
An alternative approach to full protocol-level migration is the use of Winternitz vaults, which provide application-level post-quantum protection for selected assets without changing Solana’s base protocol.
Winternitz One-Time Signatures (WOTS) are hash-based signatures: their security relies on the one-wayness of cryptographic hash functions\,\cite{buchmann2013security}.
This makes them resistant to Shor-type attacks, since there is no elliptic-curve private key to recover.
In Solana, this construction can be implemented through vault programs that require a valid Winternitz signature before funds can be moved.
Blueshift’s Solana Winternitz Vault demonstrates this model: each deposit is secured by a fresh Winternitz keypair, the public key is hashed to derive a Program Derived Address (PDA), and spending closes the current vault while creating a new one for any remaining funds\,\cite{blueshift}. 

However, they only protect assets that have been moved into the vault and do not secure the broader Solana account model, validator identities, consensus votes, fee-payer accounts, networking signatures, or arbitrary programs that continue to rely on elliptic-curve cryptography.
For this reason, Winternitz vaults are useful as an early defensive layer and emergency protection tool, while long-term quantum safety still requires native post-quantum signatures, address migration, and consensus-layer upgrades. 

\subsection{Algorand approach}
Algorand’s current PQC approach is centered on Falcon\,\cite{algorand}.
The platform already uses Falcon-1024 in state proofs, providing quantum-resistant attestations of blockchain state every 256 rounds.
Algorand has also demonstrated Falcon-secured transactions by embedding Falcon public keys into stateless LogicSig programs that verify signatures over transaction IDs.
This allows users to create Falcon-controlled accounts that can be funded and spent from like ordinary Algorand accounts, while preserving compatibility with the existing transaction pipeline.
However, this protects only selected accounts and state proofs; consensus signatures still rely on classical cryptography.

Algorand’s roadmap moves from this experimental LogicSig model toward native post-quantum accounts\,\cite{algorand_roadmap}.
The planned Q3 2026 protocol release introduces native PQ account support while preserving 32-byte addresses by deriving them from hashes of post-quantum public keys.
The roadmap also includes SDK, wallet, and AlgoKit support, followed by hybrid accounts, mixed-scheme multisig, and broader cryptographic agility. 
The remaining long-term challenge is upgrading consensus-layer components, especially Ed25519 signatures and ECC-based VRFs.

\subsection{Ethereum PQC proposals}
Ethereum’s main Falcon-integration proposal is EIP-8052, which introduces precompiled contracts for verifying Falcon-512 signatures in EVM environments\,\cite{etherium}.
The proposal splits verification into two parts: a hash-to-point step and a Falcon core verification step.
It defines both a NIST-compliant SHAKE256 version and an EVM-friendly Keccak-based version, because Keccak is already efficiently supported in the EVM. The motivation is to give Ethereum rollups and other EVM chains a practical way to adopt post-quantum signatures before quantum adversaries become practical. Falcon-512 is proposed because of its compact signature size, about 666 bytes, and efficient verification, making it more suitable for gas- and bandwidth-constrained blockchain applications than larger PQC signature schemes. However, this proposal does not replace Ethereum’s native ECDSA accounts by itself; rather, it provides infrastructure for smart contracts, rollups, wallets, and account-abstraction systems to begin verifying Falcon signatures on-chain.

\section{Conclusion}
Quantum computers invalidate the core assumption on which most existing blockchain security relies: that private keys cannot be efficiently derived from public keys.
Shor’s algorithm directly threatens ECC signature schemes, while Grover’s algorithm has a more limited effect on hash functions.
As a result, the most urgent risks are concentrated in exposed public-key accounts, reused addresses, validator keys, smart-contract administrators, stablecoin issuers, custodial wallets, and other high-value signing authorities.
The threat is especially serious because blockchain transactions are generally irreversible, and a single forged signature may be sufficient to transfer assets permanently.

Recent resource estimates suggest that progress in quantum algorithms and error correction is reducing the resources required to attack elliptic-curve cryptography\,\cite{google2026}, while the exact capabilities of future quantum systems may become visible only after attacks occur on-chain.
Therefore, the appropriate strategy is not to wait for Q-Day\,\cite{Qday_TQ}, but to begin migration while there is still time for orderly coordination.

The migration challenge differs across blockchain ecosystems.
Bitcoin benefits from hash-protected address types, but dormant and previously exposed public keys create a difficult long-term governance problem.
Ethereum and other smart-contract platforms face broader exposure through user accounts, validator signatures, stablecoins, and privileged smart-contract keys.
Solana’s public-key account model creates opportunities for immediate at-rest attacks once a CRQC exists\,\cite{anza}.
Algorand demonstrates that post-quantum secured transaction authorization is already technically feasible through Falcon-based signatures\,\cite{algorand}, although full consensus-layer migration remains a separate challenge\,\cite{algorand_roadmap}.
These examples show that there is no single migration path for all blockchains; each network must adapt PQC to its own performance, governance, and account-model constraints.

Overall, post-quantum migration should be treated not merely as a cryptographic upgrade, but as a financial stability and infrastructure resilience priority.
The growth of stablecoins, tokenized RWAs, DeFi, bridges, and custodial infrastructure will only increase the value secured by quantum-vulnerable signatures.
Fortunately, practical building blocks already exist: NIST-standardized lattice- and hash-based signatures, Falcon verification proposals for blockchain environments\,\cite{etherium}, Algorand’s Falcon-based accounts\,\cite{algorand}, Solana’s Winternitz vaults\,\cite{buchmann2013security, blueshift}, and broader research into post-quantum consensus and aggregation.
Blockchain communities, custodians, exchanges, application developers, and policymakers should begin preparing migration mechanisms before quantum attacks become practical.

\printbibliography
\end{document}